\documentclass[10pt,twocolumn,twoside]{IEEEtran}

\usepackage{graphicx}
\usepackage{cite}
\usepackage[cmex10]{amsmath}
\usepackage{amssymb}
\usepackage{mathtools}

\usepackage{tikz}
\usepackage{pgfplots}
\pgfplotsset{compat=1.3}
\usetikzlibrary{arrows,snakes,shapes}

\DeclareMathAlphabet{\mathbit}{OML}{cmr}{bx}{it}

\DeclareMathOperator{\E}{E}
\DeclareMathOperator{\T}{T}

\DeclareMathOperator{\Probability}{Pr}
\DeclareMathOperator{\Diag}{diag}

\renewcommand\vec[1]{\operatorname{vec}\left(#1\right)}
\renewcommand\arcsin[1]{\operatorname{arcsin}\left(#1\right)}

\DeclareMathOperator{\fieldR}{\mathbb{R}}

\newcommand{\diag}[1]{\Diag{\left(#1\right)}}

\newcommand{\sign}[1]{\Sign{\left(#1\right)}}

\newcommand{\ve}[1]{\boldsymbol{#1}}

\newcommand{\exdi}[2]{\E_{#1} \left[#2\right]}

\renewcommand{\exp}[1]{\operatorname{exp}\left(#1\right)}

\newcommand{\Prob}[1]{\Probability\left\{#1\right\}}

\newcommand\Sign{\operatorname{sign}}

\title{In a One-Bit Rush: Low-Latency Wireless\\Spectrum Monitoring with Binary Sensor Arrays}

\author{Manuel S. Stein and Michael Fau{\ss}\thanks{This work was supported by the German Academic Exchange Service (DAAD) with funds from the German Federal Ministry of Education and Research (BMBF) and the People Program (Marie Sk{\l}odowska-Curie Actions) of the European Union's Seventh Framework Program (FP7) under REA grant agreement no. 605728 (P.R.I.M.E. - Postdoctoral Researchers International Mobility Experience).}
\thanks{M. S. Stein is with the Mathematics Department, Vrije Universiteit Brussel, Belgium, and with the Chair for Stochastics, Universit\"at Bayreuth, Germany (e-mails: manuel.stein@vub.ac.be, manuel.stein@uni-bayreuth.de). M. Fau{\ss} is with the Signal Processing Group, TU Darmstadt, Germany (e-mail: michael.fauss@spg.tu-darmstadt.de). }
}
\begin{document}
\maketitle
\begin{abstract}
Detecting the presence of a random wireless source with minimum latency utilizing an array of radio sensors is considered. The problem is studied under the constraint that the analog-to-digital conversion at each sensor is restricted to reading the sign of the analog received signal. We formulate the resulting digital signal processing task as a sequential hypothesis test in simple form. To circumvent the intractable probabilistic model of the multivariate binary array data, a reduced model representation within the exponential family in conjunction with a log-likelihood ratio approximation is employed. This approach allows us to design a likelihood-based sequential test and to analyze its analytic performance along Wald's classical arguments. In the context of wireless spectrum monitoring for satellite-based navigation and synchronization systems, we study the achievable processing latency, characterized by the average sample number, as a function of the binary sensors in use. The practical feasibility and potential of the discussed low-complexity sensing and decision-making technology is demonstrated via simulations.
\end{abstract}
\section{Introduction}\label{sec:introduction}
In the \emph{Internet of Things} (IoT) and for upcoming standards like \emph{5th Generation Wireless Systems} (5G), instant connectivity between a massive number of wireless devices is intended. This requires to further push radio technology towards ultra-low power consumption, production cost, and circuit size while preserving the capability to quickly solve processing tasks under strict reliability requirements. 

In particular, digitalization performed by modern sensor systems has been identified as a bottleneck for ultra-low complexity \cite{Walden99,Wentzloff05, Verhelst15}. During this step, referred to as analog-to-digital (A/D) conversion, the physically measured analog sensor signals are translated to a representation which is interpretable by digital computers. As, depending on the circuit architecture, the number of A/D operations required per sample can grow exponentially $\mathcal{O}(2^b)$ with the output bits $b$, determining the appropriate amount bits required for a specific sensing application is crucial for the system design specification \cite{Wentzloff05}.

A promising approach is to reduce the A/D resolution to a single bit \cite{Ivrlac06,Mezghani12,Choi16,Gokceoglu17,Landau17,Jacobsson17} and compensate the nonlinear effect by concise probabilistic modeling and powerful algorithmic concepts which rest upon results in statistical analysis, see e.g., \cite{Stein17} and references therein. Employing such an analog-to-binary (A/B) conversion scheme has various advantages. On the one hand, it reduces the digitalization complexity to its minimum by activating a single comparator per sample. On the other hand, the size of the digital sensor data is minimized enabling efficient storage, transmission, and low-level processing. For sensing systems, sampling with low A/D resolution is often well-motivated since in several applications the digital system behind the analog sensors aims at performing algorithmic tasks which, per se, do not require perfect reconstruction of the received analog signals. 

In wireless sensor networks, a processing task of specific interest is reliable decision-making \cite{Sandell81,Viswanathan97}, i.e., distinguishing between different scenarios by using noisy radio measurements. For example, the increasing dependency of critical infrastructure (e.g., financial markets, power networks, communication infrastructure) on low-power synchronization signals provided by global navigation satellite systems (GNSS), makes it inevitable to permanently monitor the associated radio frequency bands for interference \cite{Ioannides16,Gao16}. Similarly, cognitive radio communication involves detecting active primary users within the licensed wireless spectrum \cite{Axell12}. Such monitoring problems can be formulated as statistical tests, where one searches for the optimal processing rule which decides for the underlying data-generating model. In its simple form, testing is restricted to two fully specified models. Under a reliability constraint, the task then is to decide which of the models has generated the observed sensor data. When the decision is to be made with low latency, i.e., ensuring a required error level while minimizing the number of samples, the problem statement is identified as a sequential test \cite{Wald45, Fauss15}.

Note that quantized sequential detection has been studied predominantly for sensor networks, where distributed nodes forward compressed statistics of their observations to minimize the communication overhead \cite{Hashemi89, Veeravalli93, Hussain1994, Mei2008, Yilmaz12, Chaudhari12, Wang2015}. In such a setup, it is usually assumed that the detector initially has access to the exact observations and subsequently compresses them, for example, by quantizing the likelihood ratio. Here we discuss a sequential procedure with collocated antennas featuring A/B front-ends. Therefore, the detector does not have access to the exact observations and has to base its decision exclusively on hard-limited sensor data \cite{Willett95, Stein18}. For sequential tests, this case is less well studied and the focus of existing works is on the design of optimal quantizers \cite{Tantaratana77, Nguyen2006,Teng2013}. In contrast, we investigate the design and analysis of sequential tests assuming that the quantizers are fixed, symmetric, and of minimum complexity. Moreover, the sensing model before the hard-limiting step is assumed to feature correlation, which is included and exploited in the test design.

First, we specify the signal model for the considered wireless spectrum monitoring problem and outline the resulting array data model under ideal A/D conversion with $\infty$-bit output resolution. Then we review the sequential probability ratio test (SPRT) and its average sample number (ASN). Using an approximation for the log-likelihood ratio (LLR) in the exponential family, we derive the average detection latency which can be achieved with low-complexity  sensing devices featuring A/B conversion. Finally, we discuss our theoretic findings in the context of system design for reliable low-latency GNSS spectrum monitoring and verify the practical feasibility of the presented radio technology via Monte-Carlo simulations.

\section{Problem Formulation}\label{sec:problem:form}
\subsection{System Model - Ideal Receiver}
We assume a single narrow-band wireless signal of unknown structure, impinging on a uniform linear array (ULA) with $S$ antennas
\begin{align}\label{receive:model:unquantized}
\ve{y}=\gamma\ve{A}\ve{x}+\ve{\eta}.
\end{align}
The array elements are placed at half carrier-wavelength distance. The radio front-end of each sensor exhibits two real-valued analog outputs (in-phase and quadrature), such that a digitized array snapshot can be summarized by the vector $\ve{y}\in\fieldR^M$, $M=2S$, where 
\begin{align}
\ve{y}=\begin{bmatrix}\ve{y}^{\T}_\text{I} &\ve{y}^{\T}_\text{Q} \end{bmatrix}^{\T}
\end{align}
with $\ve{y}_\text{I},\ve{y}_\text{Q}\in\fieldR^S$. The wireless source $\ve{x}\in\fieldR^2$ is composed of two independent Gaussian random variables (in-phase and quadrature)
\begin{align}
\ve{x}=\begin{bmatrix}x_\text{I} &x_\text{Q} \end{bmatrix}^{\T},
\end{align}
which are here assumed to feature
\begin{equation}
\ve{\mu}_{\ve{x}} =\exdi{\ve{x}}{\ve{x}}=\ve{0}
\quad \text{and} \quad
\ve{R}_{\ve{x}} =\exdi{\ve{x}}{\ve{x}\ve{x}^{\T}}=\ve{I},
\end{equation}
where $\exdi{\ve{u}}{\cdot}$ denotes the expectation with respect to the distribution of the random variable $\ve{u}$ and $\ve{I}$ symbolizes the identity matrix. For a narrow-band wireless signal arriving under the angle $\zeta\in\fieldR$, the array response of the ULA is characterized by the steering matrix
\begin{align}
\ve{A}=\begin{bmatrix}\ve{A}^{\T}_\text{I} &\ve{A}^{\T}_\text{Q} \end{bmatrix}^{\T},
\end{align}
where the sub-matrices $\ve{A}_\text{I}, \ve{A}_\text{Q}\in\fieldR^{S \times 2}$ have the structure \cite{SteinWSA16}
\begin{align}
\ve{A}_\text{I}=\begin{bmatrix}
\cos{\big(0\big)} &\sin{\big(0\big)}\\ 
\cos{\big(\pi\sin{(\zeta)}\big)} &\sin{\big(\pi\sin{(\zeta)}\big)}\\ 
\vdots &\vdots\\ 
\cos{\big((S-1)\pi\sin{(\zeta)}\big)} &\sin{\big((S-1)\pi\sin{(\zeta)}\big)}
\end{bmatrix}
\end{align}
and
\begin{align}
\ve{A}_\text{Q}=\begin{bmatrix}
-\sin{\big(0\big)} &\cos{\big(0\big)}\\ 
-\sin{\big(\pi\sin{(\zeta)}\big)} &\cos{\big(\pi\sin{(\zeta)}\big)} \\ 
\vdots &\vdots\\ 
-\sin{\big((S-1)\pi\sin{(\zeta)}\big)} &\cos{\big((S-1)\pi\sin{(\zeta)}\big)}
\end{bmatrix}.
\end{align}
The additive array sensor noise $\ve{\eta}\in\fieldR^M$ feature
\begin{equation}
  \ve{\mu}_{\ve{\eta}} =\exdi{\ve{\eta}}{\ve{\eta}}=\ve{0}
  \quad \text{and} \quad
  \ve{R}_{\ve{\eta}} =\exdi{\ve{\eta}}{\ve{\eta}\ve{\eta}^{\T}}=\ve{I},
\end{equation}
and the strength of the source is denoted $\gamma\in\fieldR$. Therefore, the array data \eqref{receive:model:unquantized} can be modeled as multivariate Gaussian
\begin{align}\label{multivariate:gauss}
\ve{y}\sim p_{\ve{y}}(\ve{y};\gamma)=\frac{\exp{-\frac{1}{2} \ve{y}^{\T} \ve{R}_{\ve{y}}^{-1}(\gamma) \ve{y}}}{ \sqrt{(2\pi)^{M} \det{(\ve{R}_{\ve{y}}(\gamma))}} },
\end{align}
uniquely specified by its covariance matrix
\begin{align}
\ve{R}_{\ve{y}}(\gamma)&=\exdi{\ve{y};\gamma}{\ve{y}\ve{y}^{\T}}=\gamma^2\ve{A}\ve{A}^{\T}+\ve{I}.
\end{align}
The analog radio front-end and the digitization process at each sensor are assumed to be adjusted such that the receiver observes temporally independent array snapshots, i.e., ideal analog low-pass filters of two-sided bandwidth $B$ and a sampling rate of $f_s=B$. We denote the samples of the $n$-th array snapshot by $\ve{y}_n\in\fieldR^M$ and summarize all digital data gathered up to this time instance by
\begin{align}\label{quantized:data}
\ve{Y}_n=\begin{bmatrix} \ve{y}_1 &\ve{y}_2 &\ldots &\ve{y}_n\end{bmatrix}\in\fieldR^{M\times n}.
\end{align}
\subsection{Processing Task - Sequential Signal Detection}
After the $n$-th observation, the signal processing task is to decide which of the two possible models (or hypotheses)
\begin{equation}\label{receive:hypotheses}
\mathcal{H}_0\colon \ve{y}\sim p_{\ve{y}}(\ve{y};\gamma_0)
\quad \text{or} \quad
\mathcal{H}_1\colon \ve{y}\sim p_{\ve{y}}(\ve{y};\gamma_1)
\end{equation}
is the true data-generating probability law. If the decision cannot be made while ensuring the required reliability level $\ve{\alpha}=\begin{bmatrix} \alpha_1 &\alpha_2\end{bmatrix}^{\T}$, the receiver takes an additional sample to perform the test with $n+1$ observations. Note, that the specified reliability level $\ve{\alpha}$ restricts
\begin{align}
\Prob{\text{decision in favor of }\mathcal{H}_0 | \mathcal{H}_1 } &\leq \alpha_1,\\
\Prob{\text{decision in favor of }\mathcal{H}_1 | \mathcal{H}_0 } &\leq \alpha_2.
\end{align}
The sampling instant, after which the decision process is terminated, is denoted by $n_{\text{D}}$ and the performance of the test with observations drawn from $p_{\ve{y}}(\ve{y};\gamma)$ is characterized by the ASN defined by
\begin{align}
\text{ASN}(\gamma)=\exdi{n_{\text{D}};\gamma}{n_{\text{D}}}.
\end{align}
\subsection{The SPRT for Generic Data Model}
A possible approach to construct a sequential detection procedure is the sequential probability ratio test (SPRT) discussed in \cite{Wald45}. Given the data $\ve{Y}_n$, the log-likelihood ratio (LLR)
\begin{align}
l(\ve{Y}_n) = \sum_{i=1}^n l(\ve{y}_i) = \sum_{i=1}^n \ln{ \frac{p(\ve{y}_n;\gamma_1)}{p(\ve{y}_n;\gamma_0)} }
\end{align}
is calculated and compared against two thresholds. If
\begin{align}\label{sprt:thresh:low}
l(\ve{Y}_n)\leq\ln{\frac{\alpha_2}{1-\alpha_1}},
\end{align}
the test is terminated with a decision in favor of $\mathcal{H}_0$. In case 
\begin{align}\label{sprt:thresh:up}
l(\ve{Y}_n)\geq \ln{\frac{1-\alpha_2}{\alpha_1}},
\end{align}
the test is terminated with a decision in favor of the data model $\mathcal{H}_1$. Otherwise, an additional sample is taken to continue the test with $\ve{Y}_{n+1}$. The ASN of the SPRT is approximately given by \cite{Wald45}
\begin{align}
\label{ASN0}
\text{ASN}(\gamma_0) \approx \frac{ (1-\alpha_1) \ln{\frac{\alpha_2}{1-\alpha_1}} + \alpha_1\ln{\frac{1-\alpha_2}{\alpha_1}}}{ \exdi{\ve{y};\gamma_0}{ l(\ve{y}) }}
\end{align}
and
\begin{align}
\label{ASN1}
\text{ASN}(\gamma_1) \approx \frac{\alpha_2 \ln{\frac{\alpha_2}{1-\alpha_1}} + \big(1-\alpha_2\big)\ln{\frac{1-\alpha_2}{\alpha_1}}}{ \exdi{\ve{y};\gamma_1}{ l(\ve{y}) }}.
\end{align}
\section{Analog-to-Binary Receive Model}
Realizing a receiver approximately following the high-resolution model \eqref{receive:model:unquantized} requires an A/D converter (ADC) with several bits output resolution for each sensor. To reduce the sample complexity to its minimum, here we assume that exclusively the sign of the analog signal at each sensor can be converted to the digital domain. Therefore, the SPRT has to be carried out based on the observations
\begin{align}\label{system:model:sign}
\ve{z} = \sign{\ve{y}},
\end{align}
where $\sign{\cdot}$ is defined such that
\begin{align}
\left[\ve{z}\right]_i=
\begin{cases}
+1& \text{if } [\ve{y}]_i \geq 0\\
-1 & \text{if } [\ve{y}]_i < 0.
\end{cases}
\end{align}
Note, that an ADC (flash architecture) with $b$ bits requires $2^b-1$ comparators per sample, while the A/B converter (ABC) modeled by the nonlinear operation \eqref{system:model:sign} can be realized by a single comparator per sample. However, the characterization of the multivariate Bernoulli distribution at the quantizer output \eqref{system:model:sign} forms a challenge \cite{Stein18}. In general, for all $2^M$ output constellations the integral
\begin{align}\label{likelihood:quantizer}
p_{\ve{z}}(\ve{z};\gamma)=\int_{\ve{\mathcal{Y}}(\ve{z})} p_{\ve{y}}(\ve{y};\gamma) {\mathrm d}\ve{y}
\end{align}
needs to be calculated, where $\ve{\mathcal{Y}}(\ve{z}) = \left\{ \ve{y} \in \fieldR^M \,\big|\, \ve{z} = \sign{\ve{y}} \right\}$. Evaluating \eqref{likelihood:quantizer} also requires the orthant probability, for which exact expressions are only known up to $M\leq4$. So while ABCs reduce the sampling complexity, the representation complexity of the probabilistic model increases. Therefore, in the following we approximate the binary data LLR required in the SPRT by a tractable expression.
\section{LLR Approximation in Exponential Families}
The distributions \eqref{multivariate:gauss} and \eqref{likelihood:quantizer} are members of the exponential family
\begin{align}\label{replacement:exp:family}
{p}_{\ve{z}}(\ve{z};\gamma)=\exp{\ve{\beta}^{\T}(\gamma) \ve{\phi}(\ve{z}) - \lambda(\gamma)+\kappa(\ve{z})},
\end{align}
where $\ve{\beta}(\gamma)\colon \fieldR \to \fieldR^{L}$ form the natural parameters, $\ve{\phi}(\ve{z})\colon \fieldR^{M} \to\fieldR^{L}$ are the sufficient statistics, $\lambda(\gamma)\colon \fieldR \to \fieldR$ is the log-normalizer and $\kappa(\ve{z})\colon \fieldR^{M} \to\fieldR$ constitutes the carrier measure. The LLR under exponential family models \eqref{replacement:exp:family} can, in general, be written as
\begin{equation}
l(\ve{z}) =\big(\ve{\beta}(\gamma_1)-\ve{\beta}(\gamma_0)\big)^{\T}  \ve{\phi}(\ve{z}) - \big(\lambda(\gamma_1) - \lambda(\gamma_0)\big).
\end{equation}
Distributions factorizing like \eqref{replacement:exp:family} exhibit regularity, i.e.,
\begin{align}\label{exp:family:regularity}
\exdi{\ve{z};\gamma}{ \frac{ \partial \ln{ p_{\ve{z}}(\ve{z};\gamma)} }{ \partial \gamma }}=0.
\end{align}
Therefore,
\begin{align}\label{exp:family:regularity:conseq}
\frac{\partial \lambda(\gamma)}{\partial \gamma} = \frac{\partial \ve{\beta}^{\T}(\gamma)}{\partial \gamma} \ve{\mu}_{\ve{\phi}}(\gamma),
\end{align}
with
\begin{align}\label{replacement:exp:family:mean}
\ve{\mu}_{\ve{\phi}}(\gamma) = \exdi{\ve{z};\gamma}{ \ve{\phi}(\ve{z}) }.
\end{align}
Whenever $\lambda(\gamma)$ is intractable, by using a symmetric finite difference approximation around $\frac{\gamma_1+\gamma_0}{2}$ for both derivatives in \eqref{exp:family:regularity:conseq} and
\begin{align}
\ve{\tilde{\mu}}_{\ve{\phi}}&=\ve{\mu}_{\ve{\phi}}\Big(\frac{\gamma_1+\gamma_0}{2}\Big),\\
\label{natural:difference}
\ve{b}&=\ve{\beta}(\gamma_1)- \ve{\beta}(\gamma_0),
\end{align}
one obtains the log-normalizer difference approximation
\begin{align}
\lambda(\gamma_1) - \lambda(\gamma_0) \approx \ve{b}^{\T} \ve{\tilde{\mu}}_{\ve{\phi}}.
\end{align}
This allows to write the LLR in an approximate form
\begin{align}\label{lrt:approx}
l(\ve{z})&\approx \ve{b}^{\T}\big( \ve{\phi}(\ve{z}) - \ve{\tilde{\mu}}_{\ve{\phi}} \big)=\tilde{l}(\ve{z}).
\end{align}
Whenever the natural parameter vector $\ve{\beta}(\gamma)$ is unknown, with
\begin{align}\label{replacement:exp:family:variance}
\ve{R}_{\ve{\phi}}(\gamma)&=\exdi{\ve{{z}};\gamma}{\ve{\phi}(\ve{z})\ve{\phi}^{\T}(\ve{z})}-\ve{\mu}_{\ve{\phi}}(\gamma)\ve{\mu}_{\ve{\phi}}^{\T}(\gamma)
\end{align}
at hand, one can approximate the natural difference \eqref{natural:difference}
\begin{align}\label{natural:diff:approx}
\ve{b}&\approx \ve{R}_{\ve{\phi}}^{-1}(\gamma_1) \ve{\mu}_{\ve{\phi}}(\gamma_1) - \ve{R}_{\ve{\phi}}^{-1}(\gamma_0) \ve{\mu}_{\ve{\phi}}(\gamma_0)
\end{align}
by arguing that this choice maximizes the distance between the expected values of \eqref{lrt:approx} under ${p}_{\ve{z}}(\ve{z};\gamma_0)$ and ${p}_{\ve{z}}(\ve{z};\gamma_1)$ \cite{Stein18}.
\section{Sequential Tests in the Exponential Family}\label{sec:one:bit:sprt}
\subsection{Approximate SPRT - Exponential Family Data Model}
With \eqref{lrt:approx}, an approximate SPRT for any exponential family model can be performed by comparing the approximate likelihood ratio
\begin{equation}\label{lrr:exp:approx}
\tilde{l}(\ve{Z}_n) = \sum_{i=1}^{n} \ve{b}^{\T}\big( \ve{\phi}(\ve{z}_n) - \ve{\tilde{\mu}}_{\ve{\phi}} \big) = n \ve{b}^{\T} \big( \ve{\hat{\phi}}_{\ve{z}}^{(n)} - \ve{\tilde{\mu}}_{\ve{\phi}} \big),
\end{equation}
with
\begin{align}
\ve{\hat{\phi}}_{\ve{z}}^{(n)}=\frac{1}{n} \sum_{i=1}^{n} \ve{\phi}(\ve{z}_i),
\end{align}
to the two thresholds \eqref{sprt:thresh:low} and \eqref{sprt:thresh:up}. Moreover, the ASN of such a sequential test can be approximated along \eqref{ASN0} and \eqref{ASN1} by using
\begin{align}
\label{sprt:performance:exp:approx}
\exdi{\ve{z};\gamma_i}{ l(\ve{z}) }
    &= \ve{b}^{\T} \ve{\mu}_{\ve{\phi}}(\gamma_i) - \big(\lambda(\gamma_1) - \lambda(\gamma_0)\big)\notag\\
    &\approx \ve{b}^{\T}\big( \ve{\mu}_{\ve{\phi}}(\gamma_i) - \ve{\tilde{\mu}}_{\ve{\phi}} \big)=\exdi{\ve{z};\gamma_i}{ \tilde{l}(\ve{z}) }.
\end{align}
\subsection{Approximate SPRT - Multivariate Bernoulli Data Model}
As the number of sufficient statistics for the multivariate Bernoulli distribution \eqref{likelihood:quantizer} scales $\mathcal{O}(2^M)$, we base our SPRT analysis on a pessimistic replacement model, also residing within the exponential family \cite{Stein18}. This model holds a reduced set of sufficient statistics 
\begin{align}\label{aux:statistics:quantizer}
\ve{\phi}(\ve{z})&=\ve{\Phi}\vec{\ve{z} \ve{z}^{\T}},
\end{align}
with $\ve{\Phi}$ being an elimination matrix canceling duplicate and constant diagonal statistics on $\ve{z} \ve{z}^{\T}$. The replacement with \eqref{aux:statistics:quantizer} is equivalent to the exact binary model \eqref{likelihood:quantizer} in the sense that it exhibits the same mean \eqref{replacement:exp:family:mean} and covariance \eqref{replacement:exp:family:variance} on the reduced set of statistics \eqref{aux:statistics:quantizer}. Therefore, the achievable ASN with hard-limited multivariate Gaussian observations \eqref{system:model:sign} can be calculated using \eqref{sprt:performance:exp:approx} and \eqref{natural:diff:approx}, where
\begin{align}
\ve{\mu}_{\ve{\phi}}(\gamma)&=\ve{\Phi}\vec{\ve{R}_{\ve{z}}(\gamma)}
\end{align}
is obtained via the classical arcsine law \cite[pp. 284]{Thomas69},
\begin{align}\label{covariance:quantized}
\ve{R}_{\ve{z}}(\ve{\theta})&=\frac{2}{\pi} \arcsin{ \ve{\Sigma}_{\ve{y}}(\ve{\theta}) },\notag\\
\ve{\Sigma}_{\ve{y}}(\ve{\theta})&=\diag{\ve{R}_{\ve{y}}(\ve{\theta})}^{-\frac{1}{2}} \ve{R}_{\ve{y}}(\ve{\theta}) \diag{\ve{R}_{\ve{y}}(\ve{\theta})}^{-\frac{1}{2}}
\end{align}
and the evaluation of \eqref{replacement:exp:family:variance} is accomplished \cite{SteinWSA16} via the joint application of the arcsine law and the quadrivariate orthant probability \cite{Sinn11}.

\section{Numerical Results \& Simulations}\label{sec:results}
We discuss application of the approximate SPRT \eqref{lrr:exp:approx} with the replacement \eqref{aux:statistics:quantizer} in the context of low-latency GNSS spectrum monitoring with sensor arrays featuring A/B conversion. The task is to quickly detect interference from direction $\zeta$ on the radio frequency $1.57$ GHz. At each sensor, the ideal analog pre-filter features a two-sided bandwidth of $B=2.046$ MHz. To separate radio interference from weak GNSS multi-path propagation, we set $\gamma_0=-24$ dB and $\gamma_1=-18$ dB. Further, we define $\alpha=\alpha_1=\alpha_2$ and always consider $\zeta=15^{\circ}$. 
\pgfplotsset{legend style={rounded corners=2pt,nodes=right}}
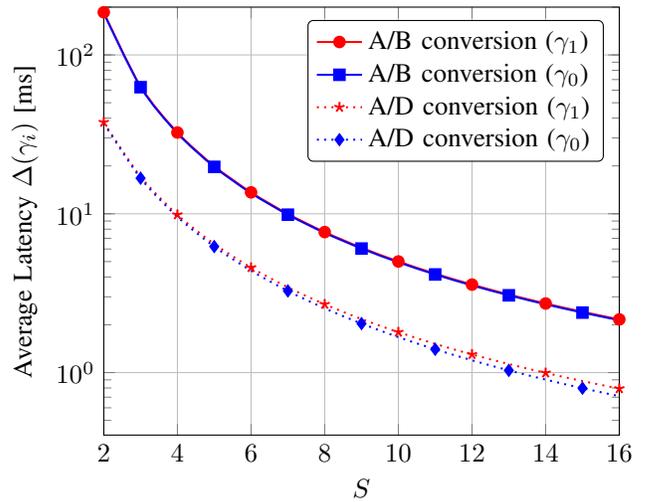
\begin{figure}[!h]
\centering
\begin{tikzpicture}[scale=1]

  	\begin{semilogyaxis}[ylabel=$\text{Average Latency $\Delta(\gamma_i)$ [ms]}$,
  			xlabel=$S$ ,
			grid,
			ymin=0,
			ymax=200,
			xmin=2,
			xmax=16,
			legend pos=north east]
			
			\addplot[red, style=solid, line width=0.75pt,smooth,every mark/.append style={solid}, mark=otimes*, mark repeat=2, mark phase=1] table[x index=0, y index=2]{ASN_NumberOfSensors_SNR0_-24_SNR1_-18_zeta15_reliability9_ANS.txt};
			\addlegendentry{A/B conversion ($\gamma_1$)}

			\addplot[blue, style=solid, line width=0.75pt,smooth,every mark/.append style={solid}, mark=square*, mark repeat=2, mark phase=2] table[x index=0, y index=1]{ASN_NumberOfSensors_SNR0_-24_SNR1_-18_zeta15_reliability9_ANS.txt};
			\addlegendentry{A/B conversion ($\gamma_0$)}
			
		 	\addplot[red, style=dotted, line width=0.75pt,smooth,every mark/.append style={solid}, mark=star, mark repeat=2, mark phase=1] table[x index=0, y index=4]{ASN_NumberOfSensors_SNR0_-24_SNR1_-18_zeta15_reliability9_ANS.txt};
			\addlegendentry{A/D conversion ($\gamma_1$)}

			\addplot[blue, style=dotted, line width=0.75pt,smooth,every mark/.append style={solid}, mark=diamond*, mark repeat=2, mark phase=2] table[x index=0, y index=3]{ASN_NumberOfSensors_SNR0_-24_SNR1_-18_zeta15_reliability9_ANS.txt};
			\addlegendentry{A/D conversion ($\gamma_0$)}
						
	\end{semilogyaxis}
	
\end{tikzpicture}
\caption{Average Latency vs. Number of Sensors ($\alpha=10^{-9}$)}
\label{fig:asn}
\end{figure}
Fig.~\ref{fig:asn} visualizes the average detection latency $\Delta(\gamma_i)=T_{\text{S}} \cdot \text{ASN}(\gamma_i)$ with A/B and ideal A/D conversion ($\infty$-bits, exact LLR). Results show that doubling the number of sensors will result in a low-complexity sensor system with competitive latency.

We define the efficiency of the low-resolution system in relation to a high-resolution system (exact LLR) with the same array size by
\begin{align}\label{def:efficiency}
\chi(\gamma_i)=\frac{\text{ASN}^{\infty\text{-bit}}(\gamma_i)}{\text{ASN}^{1\text{-bit}}(\gamma_i)}.
\end{align}
Note that the system quality measure \eqref{def:efficiency} is independent of the reliability level $\alpha$.
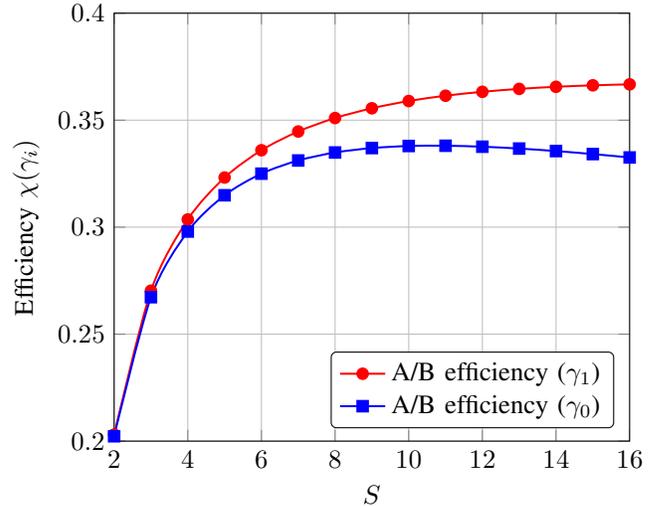
\begin{figure}[!h]
\centering
\begin{tikzpicture}[scale=1]

  	\begin{axis}[ylabel=$\text{Efficiency $\chi(\gamma_i)$}$,
  			xlabel=$S$ ,
			grid,
			ymin=0.2,
			ymax=0.4,
			xmin=2,
			xmax=16,
			legend pos=south east]
			
			\addplot[red, style=solid, line width=0.75pt,smooth,every mark/.append style={solid}, mark=otimes*, mark repeat=1] table[x index=0, y index=2]{ASN_NumberOfSensors_SNR0_-24_SNR1_-18_zeta15_reliability9_EFF.txt};
			\addlegendentry{A/B efficiency ($\gamma_1$)}

			\addplot[blue, style=solid, line width=0.75pt,smooth,every mark/.append style={solid}, mark=square*, mark repeat=1] table[x index=0, y index=1]{ASN_NumberOfSensors_SNR0_-24_SNR1_-18_zeta15_reliability9_EFF.txt};
			\addlegendentry{A/B efficiency ($\gamma_0$)}

	\end{axis}
	
\end{tikzpicture}
\caption{Efficiency vs. Number of Sensors}
\label{fig:eff}
\end{figure}
Fig.~\ref{fig:eff} depicts the analytic quality results as a function of the sensors in use. It can be seen that under interference an efficiency above $\chi(\gamma_1)=0.35$ is obtained with $S \geq 8$. For the interference-free case, the performance attains its maximum $\chi(\gamma_1)=0.33$ with $S=10$ and decreases moderatly for larger arrays. However, note that the interference-free relative latency $\chi(\gamma_0)$ is less critical than the efficiency $\chi(\gamma_1)$ for the case where a radio interferer jeopardizes the integrity of near-by GNSS receivers.
\begin{figure}[!h]
\centering
\begin{tikzpicture}[scale=1]

  	\begin{axis}[ylabel=$\text{Average LLR Value}$,
  			xlabel=$\text{Observation Duration [ms]}$ ,
			grid,
			ymin=-10,
			ymax=10,
			xmin=0,
			xmax=1.4,
			legend style={at={(0.35,0.5)},anchor=west}]
			
			\addplot[red, style=solid, line width=0.75pt,smooth,each nth point={100}] table[x index=0, y index=2]{LLR_Simulation_S_16_SNR0_-24_SNR1_-18_zeta15_reliability3_1bit.txt};
			\addlegendentry{$1$-bit sim. ($S=16, \gamma_1$)}

			\addplot[blue, style=solid, line width=0.75pt,smooth,each nth point={100}] table[x index=0, y index=1]{LLR_Simulation_S_16_SNR0_-24_SNR1_-18_zeta15_reliability3_1bit.txt};
			\addlegendentry{$1$-bit sim. ($S=16, \gamma_0$)}
			
			\addplot[green, style=solid, line width=0.75pt,smooth,each nth point={100}] table[x index=0, y index=2]{LLR_Simulation_S_8_SNR0_-24_SNR1_-18_zeta15_reliability3_ideal.txt};
			\addlegendentry{ideal sim. ($S=8, \gamma_1$)}

			\addplot[magenta, style=solid, line width=0.75pt,smooth,each nth point={100}] table[x index=0, y index=1]{LLR_Simulation_S_8_SNR0_-24_SNR1_-18_zeta15_reliability3_ideal.txt};
			\addlegendentry{ideal sim. ($S=8, \gamma_0$)}
			
		   \addplot[black, style=solid, line width=0.75pt,smooth,each nth point={100}] table[x index=0, y index=5]{LLR_Simulation_S_8_SNR0_-24_SNR1_-18_zeta15_reliability3_ideal.txt};
			
		      \addplot[black, style=solid, line width=0.75pt,smooth,each nth point={100}] table[x index=0, y index=6]{LLR_Simulation_S_8_SNR0_-24_SNR1_-18_zeta15_reliability3_ideal.txt};
		      
		      \addplot[red, style=dotted, line width=0.75pt,smooth,each nth point={100}] table[x index=0, y index=4]{LLR_Simulation_S_16_SNR0_-24_SNR1_-18_zeta15_reliability3_1bit.txt};

			\addplot[blue, style=dotted, line width=0.75pt,smooth,each nth point={100}] table[x index=0, y index=3]{LLR_Simulation_S_16_SNR0_-24_SNR1_-18_zeta15_reliability3_1bit.txt};
			
			\addplot[green, style=dotted, line width=0.75pt,smooth,each nth point={100}] table[x index=0, y index=4]{LLR_Simulation_S_8_SNR0_-24_SNR1_-18_zeta15_reliability3_ideal.txt};

			\addplot[magenta, style=dotted, line width=0.75pt,smooth,each nth point={100}] table[x index=0, y index=3]{LLR_Simulation_S_8_SNR0_-24_SNR1_-18_zeta15_reliability3_ideal.txt};
									
	\end{axis}
	
\end{tikzpicture}
\caption{LLR - Simulation vs. Analytic Result ($\alpha=10^{-3}$)}
\label{fig:sim}
\end{figure}
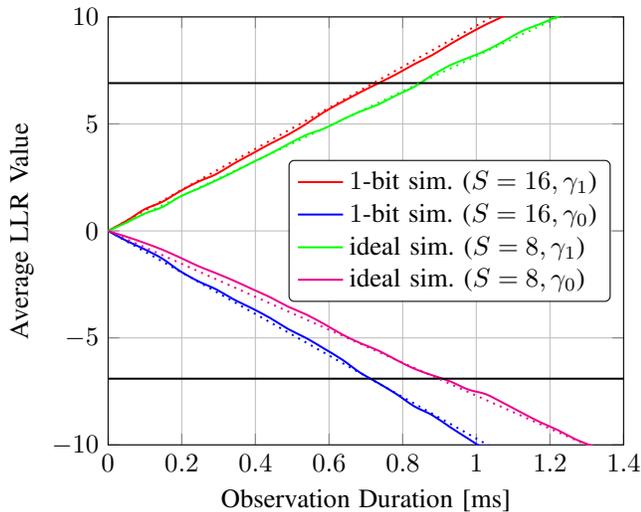
Finally, we verify the analytic ASNs \eqref{ASN0} and \eqref{ASN1} with the approximation \eqref{sprt:performance:exp:approx} by simulating the sequential test \eqref{lrr:exp:approx}. Fig.~\ref{fig:sim} shows the average log-likelihood value \eqref{lrr:exp:approx} of a low-resolution sensor system with $S=16$ and of a high-resolution version (exact LLR) with $S=8$ calculated from 200 independent test runs. The expected log-likelihood values \eqref{sprt:performance:exp:approx}, as function of the test duration, are plotted by dotted lines. We also depict the decision thresholds for $\alpha=10^{-3}$. The simulations show good correspondence between the analytic results and the simulated behavior of the approximated binary SPRT.
\section{Conclusion}
We have discussed the statistical processing problem of low-latency random signal detection with measurements obtained by an array of binary radio sensors. To circumvent the intractability of the $1$-bit model likelihood, we have employed an approximate version of the LLR which is generically valid within exponential families. The approximation enables designing a powerful sequential test along classical arguments and studying its performance analytically. In the application context of GNSS spectrum monitoring, our results show that low-cost sensing systems are capable of protecting critical infrastructure under strict latency and reliability criteria. In comparison to conventional architectures, these binary radio systems provide competitive decision-making capabilities if the hardware, power and computing resources made available by minimizing the A/D output resolution are traded for a larger number of sensor array elements.


\begin{thebibliography}{20}

\bibitem{Walden99}
R. H. Walden, ``Analog-to-digital converter survey and analysis,'' \emph{IEEE J. Sel. Areas Commun.}, vol. 17, no. 4, pp. 539--550, Apr. 1999.

\bibitem{Wentzloff05}
D. D. Wentzloff, R. Blazquez, F. S. Lee, B. P. Ginsburg, J. Powell, and A. P. Chandrakasan, ``System design considerations for ultra-wideband communication,'' \emph{IEEE Comm. Mag.}, vol. 43, no. 8, pp. 114--121, Aug. 2005.

\bibitem{Verhelst15}
M. Verhelst and A. Bahai, ``Where analog meets digital: Analog-to-information conversion and beyond,'' \emph{IEEE Solid-State Circuits Mag.}, vol. 7, no. 3, pp. 67-80, Sep. 2015.

\bibitem{Ivrlac06}
M. T. Ivrlac and J. A. Nossek, ``Challenges in coding for quantized MIMO systems,'' in IEEE Int. Symposium on Information Theory, Seattle, 2006, pp. 2114--2118.

\bibitem{Mezghani12}
A. Mezghani and J. A. Nossek, ``Capacity lower bound of MIMO channels with output quantization and correlated noise," presented at IEEE Int. Symposium on Information Theory (ISIT), Cambridge, 2012.

\bibitem{Choi16}
J. Choi, J. Mo and R. W. Heath, ``Near maximum-likelihood detector and channel estimator for uplink multiuser massive MIMO systems with one-bit ADCs,'' \emph{IEEE Trans. Commun.}, vol. 64, no. 5, pp. 2005--2018, May 2016.

\bibitem{Gokceoglu17}
A. Gokceoglu, E. Bj\"ornson, E. G. Larsson, and M. Valkama, ``Spatio-temporal waveform design for multiuser massive MIMO downlink with 1-bit receivers,'' \emph{IEEE J. Sel. Topics Signal Process.}, vol. 11, no. 2, pp. 347--362, Mar. 2017.

\bibitem{Landau17}
L. Landau, M. D\"orpinghaus, and G. P. Fettweis, ``1-bit quantization and oversampling at the receiver: Communication over bandlimited channels with noise,'' \emph{IEEE Commun. Lett.}, vol. 21, no. 5, pp. 1007-1010, May 2017.

\bibitem{Jacobsson17}
S. Jacobsson, G. Durisi, M. Coldrey, U. Gustavsson, and C. Studer, ``Throughput analysis of massive MIMO uplink with low-resolution ADCs,'' \emph{IEEE Trans. Wireless Commun.}, vol. 16, no. 6, pp. 4038--4051, Jun. 2017.

\bibitem{Stein17}
M. S. Stein, S. Bar, J. A. Nossek, and J. Tabrikian, ``Performance analysis for channel estimation with 1-bit ADC and unknown quantization threshold,'' \emph{accepted for publication in IEEE Trans. Signal Process.}, 2018, preprint: http://arxiv.org/abs/1703.02008.

\bibitem{Sandell81}
R. R. Tenney and N. R. Sandell, ``Detection with distributed sensors,'' \textit{IEEE Trans. Aerosp. Electron. Syst.}, vol. 17, no. 4, pp. 501--510, Jul. 1981.

\bibitem{Viswanathan97}
R. Viswanathan and P. K. Varshney,``Distributed detection with multiple sensors: Part I. - Fundamentals,'' \emph{Proc. IEEE}, vol. 85, no. 1, pp. 54--63, Jan. 1997.

\bibitem{Ioannides16}
R. T. Ioannides, T. Pany, and G. Gibbons, ``Known vulnerabilities of global navigation satellite systems, status, and potential mitigation techniques,'' \emph{Proc. IEEE}, vol. 104, no. 6, pp. 1174--1194, Jun. 2016.

\bibitem{Gao16}
G. X. Gao, M. Sgammini, M. Lu, and N. Kubo, ``Protecting GNSS receivers from jamming and interference,'' \emph{Proc. IEEE}, vol. 104, no. 6, pp. 1327--1338, Jun. 2016.

\bibitem{Axell12}
E. Axell, G. Leus, E. G. Larsson, and H. V. Poor, ``Spectrum sensing for cognitive radio: State-of-the-art and recent advances,'' \emph{IEEE Signal Process. Mag.}, vol. 29, no. 3, pp. 101--116, May 2012.

\bibitem{Wald45}
A. Wald, ``Sequential tests of statistical hypothesis,'' Ann. Math. Stat. vol. 16, no. 2, pp. 117-186, Jun. 1945.

\bibitem{Fauss15}
 M. Fau{\ss} and A. M. Zoubir, ``A linear programming approach to sequential hypothesis testing,'' \emph{Sequential Analysis}, vol. 34, no. 2, pp. 235--263, May 2015.
 
 \bibitem{Hashemi89}
H. R. Hashemi and I. B. Rhodes, ``Decentralized sequential detection,'' \emph{IEEE Trans. Inform. Theory}, vol. 35, no. 3, pp. 509--520, May 1989.

\bibitem{Veeravalli93}
V. V. Veeravalli, T. Basar, and H. V. Poor, ``Decentralized sequential detection with a fusion center performing the sequential test,'' \emph{IEEE Trans. Inform. Theory}, vol. 39, no. 2, pp. 433--442, Mar. 1993.

\bibitem{Hussain1994}
A. M. Hussain, ``Multisensor distributed sequential detection," \textit{IEEE Trans. Aerosp. Electron. Syst.}, vol. 30, no. 3, pp. 698--708, Jul. 1994.

\bibitem{Mei2008}
Y. Mei, ``Asymptotic optimality theory for decentralized sequential hypothesis testing in sensor networks," \textit{IEEE Trans. Inform. Theory}, vol. 54, no. 5, pp. 2072--2089, May 2008.

 \bibitem{Yilmaz12}
Y. Y{\i}lmaz, G. V. Moustakides, and X. Wang, ``Cooperative sequential spectrum sensing based on level-triggered sampling,'' \emph{IEEE Trans. Signal Process.}, vol. 60, no. 9, pp. 4509--4524, Sept. 2012.

\bibitem{Chaudhari12}
S. Chaudhari, J. Lund{\'e}n, and V. Koivunen, ``Effects of quantization and channel errors on sequential detection in cognitive radios,'' in 46th Annual Conference on Information Sciences and Systems (CISS), Princeton, 2012, pp. 1--6.

\bibitem{Wang2015}
Y. Wang and Y. Mei, ``Quantization effect on the log-likelihood ratio and its application to decentralized sequential detection," \textit{IEEE Trans. Signal Process.}, vol. 61, no. 6, pp. 1536--1543, Mar. 2015.

\bibitem{Willett95}
P. Willett and P. F. Swaszek, ``On the performance degradation from one-bit quantized detection,'' \textit{IEEE Trans. Inf. Theory}, vol. 41, no. 6, pp. 1997--2003, Nov. 1995.

\bibitem{Stein18}
M. S. Stein, ``Asymptotic signal detection rates with 1-bit array measurements,'' to be presented at IEEE Int. Conference on Acoustics, Speech and Signal Processing (ICASSP), Calgary, Canada, 2018, preprint: http://arxiv.org/abs/1711.00739.

\bibitem{Tantaratana77}
S. Tantaratana and J. Thomas, ``Quantization for sequential signal detection,'' \emph{IEEE Trans. Commun.}, vol. 25, no. 7, pp. 696--703, Jul. 1977.

\bibitem{Nguyen2006}
X. Nguyen, M. J. Wainwright, and M. I. Jordan, ``On optimal quantization rules for sequential decision problems," in IEEE International Symposium on Information Theory, Seattle, 2006. 

\bibitem{Teng2013}
D. Teng and E. Ertin, ``Optimal quantization of likelihood for low complexity sequential testing," IEEE Global Conference on Signal and Information Processing, Austin, 2013.

\bibitem{Thomas69}
J. B. Thomas,  \textit{Introduction to Statistical Communication Theory}. Hoboken, NJ: John Wiley \& Sons, 1969.

\bibitem{SteinWSA16}
M. Stein, K. Barb\'e, and J. A. Nossek, ``DOA parameter estimation with 1-bit quantization - Bounds, methods and the exponential replacement", in Int. ITG Workshop on Smart Antennas, Munich, 2016, pp. 1-6.

\bibitem{Sinn11}
M. Sinn and K. Keller, ``Covariances of zero crossings in Gaussian processes," \textit{Theory Probab. Appl.}, vol. 55, no. 3, pp. 485--504, 2011.

\end{thebibliography}
\end{document}